\documentclass[namedreferences,12pt]{solarphysics}
\usepackage[hyperref,optionalrh,solaromanenum]{spr-sola-addons}  
\usepackage{graphicx}                    
\usepackage{amssymb}                  
\usepackage{color}                         
\usepackage{breakurl}                    
\usepackage{soul}
\usepackage{hyperref}

\newcommand{\degree}{\ensuremath{^\circ}}

\newcommand{\soho}{{\em SOHO{}}}
\newcommand{\sdo}{{\em SDO{}}}
\newcommand{\stereo}{{\em STEREO{}}}

\runningtitle{Revisiting the Extended Solar Cycle}
\runningauthor{S. W. McIntosh {\em et~al.}}

\begin{document}
\begin{opening}
\title{Deciphering Solar Magnetic Activity: 140 Years Of The `Extended Solar Cycle' -- Mapping the Hale Cycle}

\author[addressref={1},corref,email={mscott@ucar.edu}]{\inits{S.W.}\fnm{Scott W.}~\lnm{McIntosh}~\orcid{0000-0002-7369-1776}}
\author[addressref={2,3}]{\inits{R.J.}\fnm{Robert J.}~\lnm{Leamon}~\orcid{0000-0002-6811-5862}}
\author[addressref={1}]{\inits{R.E.}\fnm{Ricky}~\lnm{Egeland}~\orcid{0000-0002-4996-0753}}
\author[addressref={1}]{\inits{M.D.}\fnm{Mausumi}~\lnm{Dikpati}~\orcid{0000-0002-YYYY-ZZZZ}}
\author[addressref={4}]{\inits{R.C.A.}\fnm{Richard C.}~\lnm{Altrock}~\orcid{0000-0002-YYYY-ZZZZ}}
\author[addressref={5,6}]{\inits{D.B.}\fnm{Dipankar}~\lnm{Banerjee}~\orcid{0000-0003-4653-6823}}
\author[addressref={7}]{\inits{S.C.}\fnm{Subhamoy}~\lnm{Chatterjee}~\orcid{0000-0002-YYYY-ZZZZ}}
\author[addressref={8}]{\inits{A.K.S.}\fnm{Abhishek K.}~\lnm{Srivastava}~\orcid{0000-0002-1641-1539}}
\author[addressref={9}]{\inits{M.V.}\fnm{Marco}~\lnm{Velli}~\orcid{0000-0002-YYYY-ZZZZ}}

\address[id={1}]{National Center for Atmospheric Research, High Altitude Observatory, P.O. Box 3000, Boulder, CO~80307, USA}
\address[id={2}]{University of Maryland--Baltimore County, Goddard Planetary Heliophysics Institute, Baltimore, MD 21250, USA}
\address[id={3}]{NASA Goddard Space Flight Center, Code 672, Greenbelt, MD 20771, USA}
\address[id={4}]{National Solar Observatory, 3665 Discovery Drive, Boulder, CO 80303, USA}
\address[id={5}]{Aryabhatta Research Institute of Observational Sciences (ARIES), Manora Peak, Nainital-263 002, India}
\address[id={6}]{Indian Institute of Astrophysics, Koramangala, Bangalore 560034, India}
\address[id={7}]{Southwest Research Institute, 1050 Walnut Street, Suite 300, Boulder, CO 80302, USA}
\address[id={8}]{Department of Physics, Indian Institute of Technology, Varanasi-221005, India}
\address[id={9}]{Department of Earth, Planetary \& Space Sciences, University of California Los Angeles, Los Angeles, CA 90095, USA}

\begin{abstract}
\textcolor{blue}{We investigate the occurrence of the ``extended solar cycle'' (ESC) as it occurs in a host observational data spanning 140 years. Investigating coronal, chromospheric, photospheric and interior diagnostics we develop a consistent picture of solar activity migration linked to the 22-year Hale (magnetic) cycle using superposed epoch analysis (SEA) using previously identified Hale cycle termination events as the key time for the SEA. Our analysis shows that the ESC and Hale cycle, as highlighted by the terminator-keyed SEA, is strongly recurrent throughout the entire observational record studied, some 140 years. Applying the same SEA method to the sunspot record confirms that Maunder's butterfly pattern is a subset of the underlying Hale cycle, strongly suggesting that the production of sunspots is not the fundamental feature of the Hale cycle, but the ESC is. The ESC (and Hale cycle) pattern highlights the importance of 55\degree\ latitude in the evolution, and possible production, of solar magnetism.} 
\end{abstract}

\end{opening}

\section{Introduction}

One of the greatest challenges in astrophysics lies in understanding how the Sun generates and cyclically modulates its global-scale magnetic field. For over 400 years solar observers have pondered the canonical marker of that magnetic field---the sunspot. It took more than 200 years after sunspot sketching and cataloging started before it was discovered that the number of sunspots waxes and wanes over an approximately 11 year period \citep{Schw49}. A half century later, mapping the latitudinal variation of the spotted Sun yielded the ``butterfly diagram,'' a pattern progressing from $\sim$30\degree\ latitude to the equator over the $\sim$11 year period \citep{Maun04}. In the golden age of solar astronomy---soon to become solar physics---that followed, it was first suggested and then demonstrated that sunspots were sites of intense magnetism protruding through the Sun's photosphere \citep{1908ApJ....28..315H, Hale19} and that the polarities of the butterfly's wings alternated in sign with a period of about 22 years \citep{1925ApJ....62..270H}. The 11(-ish) year periodicity of sunspot number and the 22(-ish) year periodicity of magnetic polarization must therefore be inextricably linked \citep[e.g.,][]{2010LRSP....7....1H}, but how? 

\begin{figure}[ht]
\centering
\includegraphics[width=\linewidth]{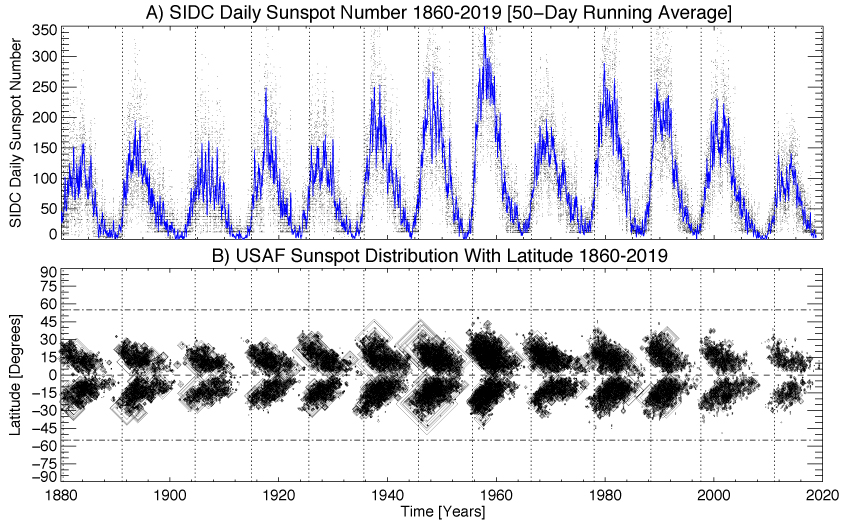}
\caption{Comparing the 140 years of sunspot evolution as visualized by the Royal Observatory of Brussels, Solar Influences Data analysis Center (SIDC) and US Air Force (USAF). Panel~A shows the daily sunspot number (fine dots) with a 50-day running average superposed in blue. Panel~B shows the latitude-time distribution of sunspots over the same time frame, the size of the symbol plotted represents the relative size of the sunspot. In each panel the vertical dotted lines are the times of the magnetic (Hale) cycle terminations \citep{2019arXiv190109083M} while the horizontal dot-dashed lines signify 55\degree\ latitude.}
\label{fig:f1}
\end{figure}

The extremely well-tended sunspot data catalog \citep[e.g.,][]{2015SpWea..13..529C}, including the limited representation shown in Figure~\ref{fig:f1}, has been scrutinized time and time again. Indeed, it has been, and will continue to be, exhaustively mined to reveal any hint of the underlying process or processes responsible for the enigmatic spots and their variation. At the largest scale, the modulation of sunspot number both in time and in time \emph{and} space, have presented primary targets for the astrophysics community with an understanding of the Sun's omnipotent magnetism as the goal. Further, the well correlated radiative analog of the Sun's magnetic variability the disk-integrated calcium indices, observed for over a century  \citep[e.g.,][]{ 1989ApJ...337..964S, Bertello:2016, Egeland:2017}, have created a means to standardize the approach of the astrophysical community to understanding solar and stellar activity en masse \citep[e.g.,][]{1978ApJ...226..379W, Baliunas:1995, Egeland:2017:thesis}, as the ability to resolve spot activity on distant stars remains in its infancy \cite[e.g.,][]{Berdyugina:2005,Morris:2017}. Any theory designed to understand the origins of the Sun's magnetism and its spotty conundrum, must replicate these measures to be plausible. When we reach our conclusion we will ask, was this dataset ever sufficient to answer the problem at hand?

The universality of the 11-year solar activity ``canon'' cannot be ignored. Simply put, all investigations of the Sun's influence on the Earth, the solar system and its other planetary bodies are intimately tied to the standard candles of “solar minimum” and “solar maximum”. These terms are used widely to describe magnetically quiet and active spells of solar activity that are tied to the dearth or glut of sunspot productivity. {\em As we will see their use is conceptually limiting, but this is not a paper to go into the myriad of ways in which the magnetic evolution of our star is more devious than face value.}

As yet, no theory can claim to replicate the underlying physics of the problem from first principles \citep{1987SoPh..110...11P} although many have tried, and some do better than others \citep[e.g.,][]{2010LRSP....7....3C}. The class of theories that have developed to explain the gross magnetic variability over the 60 years since routine observations of the Sun's global magnetism became possible \citep[e.g.,][]{1913ApJ....38...27H, 1961ApJ...133..572B} are generally grouped by the term ``dynamo theory.'' A dynamo theory tries to capture the Sun's ability to convert toroidal magnetic fields into poloidal magnetic fields and vice versa utilizing solar internal differential rotation,
turbulent convection and circulatory patterns. Four principal forms of dynamo model are prevalent: the ``Babcock-Leighton Dynamo;'' the ``Flux-Transport Dynamo''; the ``dynamo wave;'' and fully convective 3D magnetohydrodynamic (MHD) models. These concepts are beautifully and insightfully discussed in Charbonneau's review \citep{2010LRSP....7....3C} although the interested reader should read about Parker's dynamo dilemma \citep{1987SoPh..110...11P} and the original literature \citep{1955ApJ...122..293P, 1964ApJ...140.1547L, 1991ApJ...383..431W}. Of these dynamo concepts MHD simulations attempt to directly recover the plasma state of the Sun's interior, while the first two are observationally-motivated kinematic models, and the remaining concept tries to explain the magnetic progression in terms of global wave-like motions. Regardless of the {\it a~priori} assumptions made in formulating the theory these models have the common goals of replicating the sunspot number modulation, the butterfly diagram progression, and the alternating polarity of the latter. Little attention has been paid to the situation where these variations place insufficient constraints on the physical problem at hand, however. There is an incredibly vast body of literature attempting to explain these phenomena (over 1,000 refereed articles with 30,000 citations with the phrase `solar dynamo' in the title or abstract since 1970). The skeptical scientist may worry about the true size of the dynamo problem's null space when so many researchers, wielding a vast array of models, can replicate these grand metrics, but lack success when used in any forward-looking way without constant ingestion of data or adjustment of the array of ``free'' parameters lurking in the background \citep{1993ApJ...408..707P}. We will return to the dynamo dilemma.

\subsection{The Extended Solar Cycle (ESC)}
The term ``Extended Solar Cycle'' first appeared in the literature in 1988 although observational evidence had been appearing over that decade indicating that magnetic activity from one ``cycle'' overlaps for some period of time with the last \citep{1983A&A...120L...1L}, often up to a few years. The culmination of many years of painstaking observation, cataloging, and individual publication by a number of prominent observers of the time---the extended cycle is a pattern formed by a host of observables in latitude and time. Some of those observables are easily associable with magnetism in nature, prominences and filaments \citep[e.g.,][]{1933MmArc..51....5B, 1975SoPh...44..225H, 1992ASPC...27...14M, 2016SoPh..291.1115T} and ephemeral active regions \citep[e.g.,][]{1973SoPh...32..389H}, and the gross features in the Sun's green-line corona \citep[e.g.,][]{1997SoPh..170..411A} while some, like the zonal flow patterns of the torsional oscillation \citep[e.g.,][]{1980ApJ...239L..33H, 1987Natur.328..696S}, still defy clear explanation. As presented by Wilson, these multi-scale manifestations of solar activity, shared a common trend. The resulting composite pattern, simply put, followed the wings of the sunspot butterfly backward in time by up to ten years and to significantly higher latitudes such that, over a few sunspot cycles it was clear that the extended butterflies overlapped one another in space and time. The extended solar cycle, or ESC, was born. It is interesting to note that papers predating Wilson's referred to the long baseline of geomagnetic data being consistent with the signature of temporally overlapping activity cycles on the Sun \citep{1975JGR....80..111M, 1981SoPh...70..173L}. This is a matter that Wilson himself addresses later in his excellently accessible monograph on solar and stellar activity cycles \citep{1994ssac.book.....W}; however, little was done to integrate the latter into the ESC picture despite being an obvious line of research pursuit.

\subsection{The Contemporary ESC}
Wilson's largely observational paper is poorly cited [$N$=125 at time of writing] despite, or maybe due to, the clear complexity it adds to the solar activity puzzle---activity cycles clearly overlap. Compared to the powerhouse solar dynamo papers of Babcock, Leighton, Parker, and Wang \& Sheeley noted above, Wilson's paper has a tiny fraction of the citations of each over the same timeframe. Why? The spatio-temporal overlapping, going back almost a decade in time and reaching far higher latitudes, could be a problem to some of the popular dynamo theories discussed above ---as we'll discuss later. Those theories were also conceived before Wilson's declaration, but never revised. Again, why? It is the view of the authors that Wilson and collaborators ``failed'' in one crucial arena---they were unable to identify how the phasing of their extended cycle pattern could explain the modulation of sunspots, grossly in number or the classical butterfly pattern. Had they managed, this would have drawn attention to the nature of the ESC as a more ``fundamental'' mode of the system. 

\begin{figure}[!ht]
\centering
\includegraphics[width=0.85\linewidth]{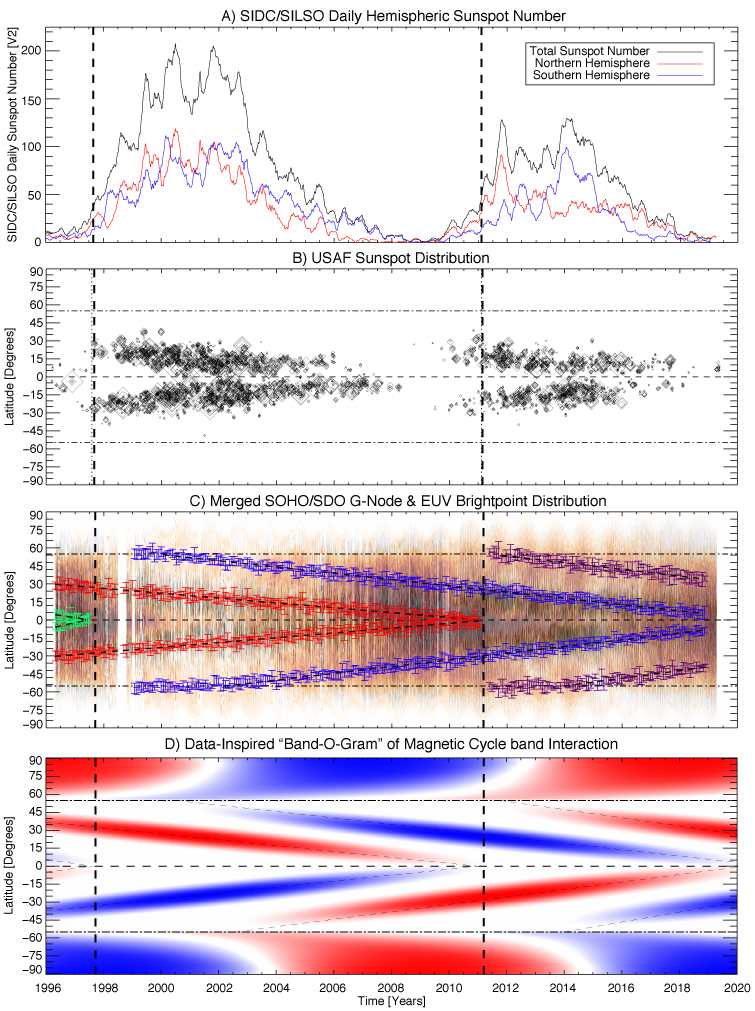}
\caption{The contemporary Extended Solar Cycle (ESC) picture as originally developed by M2014. Panel~A of the figure show the Royal Observatory of Brussels, Solar Influences Data analysis Center (SIDC) daily total (black line) and hemispheric (north \-- red; south \-- blue) sunspot number where each has a 50-day running average applied. Panel~B shows the latitude-time distribution of sunspots over the same time frame, the size of the symbol plotted represents the relative size of the sunspot. Panel~C shows a temporally extended version of EUV brightpoint density analysis and related band identification originally published in M2014 and subsequently updated \citep{2017FrASS...4....4M}. The plotted points and error bars signify the latitudinal location and extent of the EUV brightpoint density bands. The bands are colored related to the sunspot cycle they delineate and shape (22 \-- green; 23 \-- red; 24 \-- blue; 25 \-- purple). Finally, in panel~D, we show the data-motivated ``band-o-gram'' as a means to present the latitudinal evolution of the magnetic polarities of the 22-year magnetic activity cycle. In each panel the vertical dashed lines are the times of the magnetic (Hale) cycle terminations \citep{2019arXiv190109083M} while the horizontal dot-dashed lines signify 55\degree\ latitude.}
\label{fig:f2}
\end{figure}

Such was the unheralded nature of Wilson and company's pioneering work that the authors of this work were blissfully unaware of its presence until near completion of their own work in this area \citep{Mac14} (\textcolor{blue}{hereafter M2014}). They investigated the spatio-temporal occurrence of ubiquitous small features in the Sun's corona called (coronal) Bright Points (or BPs) and the magnetic scale to which they were tied in the Sun's photosphere. McIntosh and colleagues were able to demonstrate that, for the epoch covered by the Solar and Heliospheric Observatory [\soho\/] and Solar Dynamic Observatory [\sdo\/], 1996--2013, or some 17 years, these small scale magnetically-rooted coronal features traced out a significant extension of the sunspot butterfly. They were able to identify overlapping bands of oppositely polarized magnetic activity responsible for sunspot cycles 22, 23, 24 and 25.

Figure~\ref{fig:f2} compares the variation in the Sunspot number, total and hemispheric, with the Sunspot distribution with latitude and time, the EUV BP density progression and the data-inspired phenomenological concept that was constructed to demonstrate the global-scale magnetic interactions taking place as a function of time that M2014 credit with shaping the sunspot progression and amplitude. M2014 redefined the landmarks of the sunspot cycle in the context of the ``band-o-gram.'' as belonging to epochs of interaction between oppositely polarized global-scale magnetic systems.


M2014 inferred that the modulation of the Solar Cycle occurred via a process that they dubbed ``magnetic teleconnection'' \-- \textcolor{blue}{an analog to the meteorological term ``teleconnection,'' that is related to each other at large distances} \citep{Barnston1987,Trenberth}. In other words, a phenomenon where atmospheric low-frequency variability (such as planetary waves) are temporally correlated between disparate physical locations over extended periods of time, like the North Atlantic Oscillation (NAO) and the El Ni\~no Southern Oscillation (ENSO). Further, it is accepted that gravity waves form the means of communication between the various circulatory patterns and the planetary waves. Magnetic teleconnection, therefore, was proposed to describe the process by which the large-scale magnetic interactions take place between the oppositely polarized, overlapping magnetic bands of the 22-year long solar magnetic activity cycle within a solar hemisphere and across the solar Equator (see, {\em e.g.}, Figure~8 of M2014). This analogy is strengthened by the subsequent discovery that the magnetic systems display the characteristics of planetary scale Rossby waves \citep[e.g.,][]{2017NatAs...1E..86M}. It is possible that other magnetic structures in the solar interior, like the strongly twisted ``wreaths'' of magnetic fields (one per hemisphere), that are suggested by some MHD simulations \citep[e.g.,][]{2013ApJ...762...73N} could present a similar magnetic signature.


\subsection{Cycle Terminators}
The picture of overlapping activity bands, visible in Figure~\ref{fig:f2}, also shows an important element of the analysis to follow: the concept of the terminator.
Stated simply the terminator is the time at which the magnetic activity bands (those presently responsible for spots) meet at the equator and cancel one another. The timescale over which these terminator events occur, originally discussed in M2014, is remarkably short for a stellar interior \citep{2019arXiv190109083M}, at around one solar rotation. Shortly following the death of the old cycle bands, a systematic increase in magnetic flux emergence across many longitudes occurs at mid-latitudes through the growth of sunspots, and at high latitudes with the launch of the ``rush to the poles.'' 
Thus, termination events mark the end of one magnetic (and sunspot) cycle at the solar equator and the start of the next solar cycle at mid-latitudes. They also mark the start of the rush to the poles, and the polar reversal process at high latitudes. Establishing the temporal separation of many such events would permit us to estimate the speed at which information around the change of state is carried through the medium \-- providing physical insight into coupling of the activity bands and the solar interior itself.

In Figure~\ref{fig:f2} the two terminators in the \soho{} era are shown as thick vertical dashed lines, one in late 1997 and another in early 2011. These terminators mark the transition from one sunspot cycle into the next while also marking the end of a magnetic activity cycle. As such they are not just arbitrary times, like a sunspot/solar minimum or (total) sunspot/solar maximum, in the phasing of solar activity---sunspot minimum is a time when four oppositely polarized magnetic bands surround the equator and negate (partially cancel) one another while sunspot maximum is a time when flux emergence begins to be throttled back by the introduction of additional magnetized bands present per hemisphere. In what follows we will use the start/end nature of the terminators as a new fiducial clock to explore solar activity. 

\section{Observations}
In the following sections we use a host of observations and previously published data analysis methods to show that the extended solar cycle can be identified and traced back for at least some 140 years. We analyze observations and catalogs of filaments from several observatories (1880--Present: Arcetri, Meudon, and Kislovodsk observatories, to name only three), and the coronal ``Green Line'' intensity scans collected  by the NOAA National Centers for Environmental Information (NCEI)  (1939--2009) from a host of observatories around the globe. We compare the latter to contemporary pictures of the coronal intensity scan available from the fleet of EUV imagers in operation: the Solar and Heliospheric Observatory ({\em SoHO}) Extreme-ultraviolet Imaging Telescope \citep[EIT; ][]{1995SoPh..162..291D}; the Solar Terrestrial Relations Observatory ({\em STEREO}) Extreme-UltraViolet Imager \citep[EUVI;][]{2008SSRv..136...67H}; and the Solar Dynamics Observatory (\sdo) Atmospheric Imaging Array \citep[AIA; ][]{2012SoPh..275...17L}. Beginning with the contemporary datasets we look for analogs for the patterns of global-scale flux emergence and migration discussed in Figure~\ref{fig:f2}.

\begin{figure}[ht]
\centering
\includegraphics[width=1.0\linewidth]{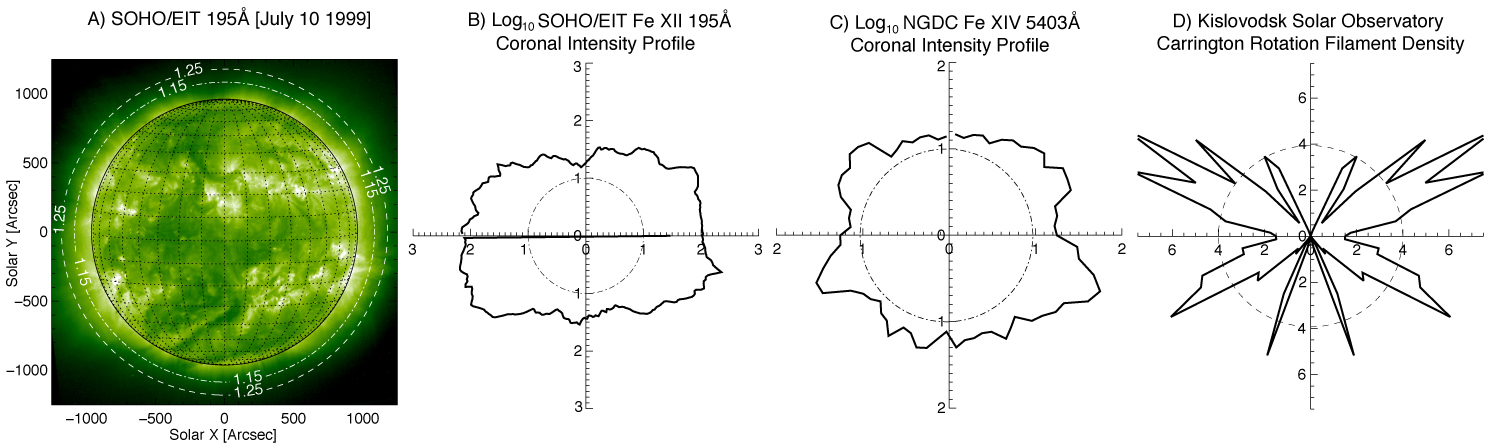}
\caption{ \protect\citep[Adapted from][]{1997SoPh..170..411A} Comparing histogram of EUV Intensity (A) from July 10, 1999 with the derived annular coronal intensity from \soho (B; EIT) and the NGDC Coronal Green Line archive (C), with the density of filaments as a function of position angle.}
\label{fig:fA}
\end{figure}

\subsection{Methods}
 To begin we provide the reader with insight into the straightforward data analysis methods employed for the extended temporal identification and discussion of the Extended Solar Cycle. 

\subsubsection{Filament Density}
Following the original analysis presented by Lockyer \citep{1903MNRAS..63..481L,1931MNRAS..91..797L} and subsequent reports by \cite{1933MmSAI...6..479B} of the Arcetri filament observations, the evolution of the latitudinal population density of chromospheric filaments was diligently monitored over many decades. These records are readily comparable to those of contemporary investigators \citep[e.g.,][]{1993JGR....9813177H, 2014SSRv..186..169C,2016SoPh..291.1115T}.

The first step in this analysis is to form the filament (when seen on the disk, or prominence, when seen on the limb) population density histogram as a function of latitude over the relevant timescale. Early investigations used and published annual statistics \citep{1933MmSAI...6..479B}, but as time progressed the methodology was applied to the shorter rotational timescales. Once these population density histograms are constructed, the authors identify the latitudinal maxima in the dataset. These population density maxima are readily tracked in time and examples can be seen in \cite{1931MNRAS..91..797L} (Figure~1\--2) and \cite{1933MmSAI...6..479B} (Figs.~1\--52). Very similar methods are used to monitor the centroids and movement of EUV BP density histograms that permit the construction of the colored bands in Figure~\ref{fig:f2}.

\subsubsection{Coronal Intensity}
%

The 5303\AA{} transition between the ground-state fine-structure levels of 
Fe~{\sc xiv} gives rise to the strongest forbidden line in the coronal spectrum and the brightest of all coronal emission lines in the visible. This is the so-called ``coronal green line.'' 
Following the work of Wilson, Altrock and others highlighted earlier \citep[e.g.,][]{1987SoPh..110....1W, 1997SoPh..170..411A, Wil88, 1994ssac.book.....W}, one of the principal mechanisms by which the ESC is/was monitored was via the latitudinal evolution of green line intensity maxima. 
The bulk of these daily measurements were made using coronagraph observations \citep{1930BuAst...6..305L} and a dedicated photometric method \citep{1974SoPh...36..343F}, or similar. The connection between the morphology of the global-scale corona and the passage of prominences (and filaments) has been of little debate for over one hundred years \citep{1903MNRAS..63..481L}. 

The daily record of green line coronal scans span 1939 to 2009. The scans can be found at the website of National Centers for Environmental Information\footnote{\url{https://www.ngdc.noaa.gov/stp/solar/corona.html}. The green line data at NOAA/NCEI were provided by Dr. V. Rusin of the Astronomical Institute, Slovak Academy of Sciences. Several coronagraphic stations were used in the database construction, with Lomnicky Peak being the primary station since 1965 \citep{1992sers.conf..168R}.}. The archived coronal brightness scans were achieved by stepping the photometer in 5\degree{} steps clockwise around the limb, at a distance of 1.15 solar radii with a sampling width of 0.05 solar radii. The values provided in millionths of intensity of the solar disk. The panels of Figure~\ref{fig:fA}, adapted from Figs.~1 and~2 of \cite{1997SoPh..170..411A}, illustrates coronal scans representative of solar minimum (left) and maximum (right) conditions in 1985. In much the similar way to the maximum identification applied to the BP and filament density distributions we identify the locations in latitude of the intensity distribution maxima and chart their position at each timestep. Over time this results in a maximum density map like in Figure~2 of \cite{1988sscd.conf..414A}.   

Contemporary space-borne broadband EUV observations permit coronal intensity scans to be constructed in a fashion similar to the historical emission line coronagraphic scans of the previous section. We have chosen the EUV passband of 195\AA{} (193\AA{} in \sdo/AIA) to replicate and extend the coronal intensity scan record, and its subsequent analysis, into the present day. Constructing a mask over the EUV imaging data at a distance of 1.15 solar radii with a width of 0.05 solar radii we then construct the intensity scan that approximates the evolution of Fe~XII emission in the corona. Given the improved spatial resolution of the space-borne EUV imaging data over the green line coronagraphic data, the EUV diagnostics are extracted with a 1\degree\ resolution around the limb.

\subsubsection{Processing Latitudinal Evolution}
With daily, monthly, or annual measures of filament density, and/or coronal intensity maxima at hand we then begin to explore long-term systematic patterns that may develop. Depending on the initial timescale of the data under consideration we perform a windowed average over a suitable period of time, over a solar rotation for daily data, or over a year for monthly data where we can aggregate maxima and create histograms of maxima density versus latitude. Using the same maximum detection algorithm as above we identify the local maxima of the density histograms as a function of solar latitude.

\section{Historical Picture}
In the following section we will study the evolution of solar features dating back around 140 years. The filament density record, obtained from published reports delivered by a host of observatories around the globe, span from the dawn of solar H$\alpha$ photography ($\sim$1870) to the present. The coronal Fe~{\sc xiv} 5303\AA{} ``green line'' archive covers seventy years from 1939 to 2009. In analyzing these data we will follow the simple analysis steps presented above.

\begin{figure}[!ht]
\centering
\includegraphics[width=\linewidth]{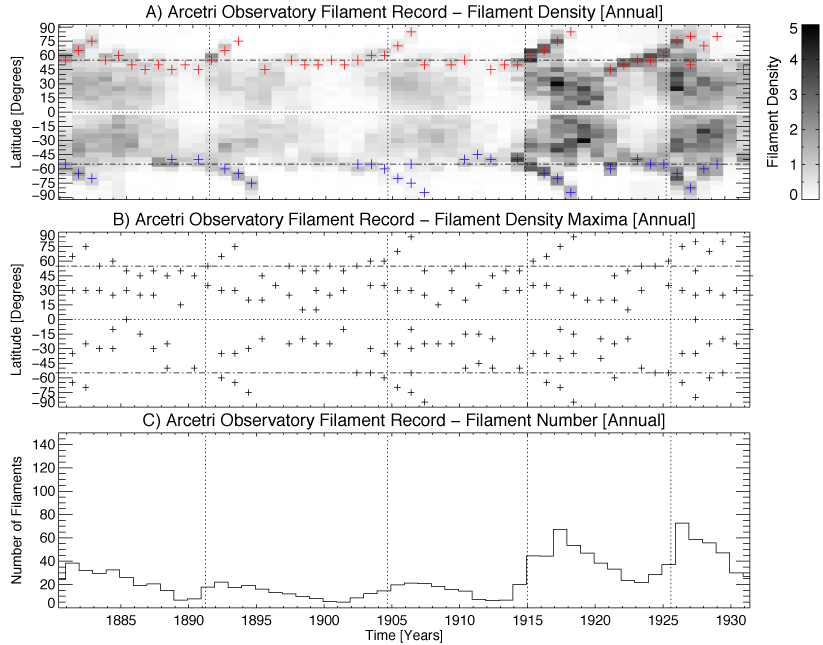}
\caption{Evolution of disk filaments on an annual timescale, 1860--1932, provided by the Arcetri Astrophysical Observatory \citep{1933MmArc..51....5B}. From top to bottom we show the annual filament density (in 5\degree\ latitude bins) versus latitude. Included in the panel are the positions of the highest latitude filament in the northern (red $+$ symbol) and southern (blue $+$ symbol) hemisphere. The black $+$ symbols of the middle panel indicate the latitude of the annual filament density maxima. The lower panel shows the annual average number of filaments. In each panel the vertical dotted lines are the times of the magnetic (Hale) cycle terminations \citep{2019arXiv190109083M}.}
\label{fig:f3}
\end{figure}

\subsection{Solar Filaments}
\textcolor{blue}{First published in 1933 by \cite{1933MmArc..51....5B} using photographic observations of the solar disk in the Balmer $\alpha$ line of Hydrogen from the Arcetri Astrophysical Observatory.} Bocchino and colleagues measured and catalogued the latitude of clearly identifiable filaments above the solar limb. They presented their data in tables of annual filament density, see Figure~\ref{fig:f3}. Beyond the butterfly-like structure in filament density at lower latitudes, we (like the earlier observers) note the clear quasi-periodic excursions to the poles in the upper and central panels of Figure~\ref{fig:f3}. We will return to their identification and analysis below.

Interestingly, practically identical patterns were published in contemporary work by \cite{1931MNRAS..91..797L} although Bocchino was the first to publish the tables in addition to graphics. We should also note that Lockyer was extending an earlier analysis, from nearly three decades earlier by the same author \citep{1903MNRAS..63..481L} where the first three decades of observation from the Kew Observatory---using technology that was shared with colleagues in Arcetri---showed patterns in filament (and prominence) evolution \citep{1892MmSSI..20..135R}. Bocchino and Lockyer independently verified each others' analyses---the latter formed a connection to the morphology of the corona at total solar eclipses that we will appeal to in the very next section. The clear excursions of the highest latitude filaments from a perch around 55\degree\ latitude were noted by all early authors. 

\begin{figure}[!ht]
\centering
\includegraphics[width=\linewidth]{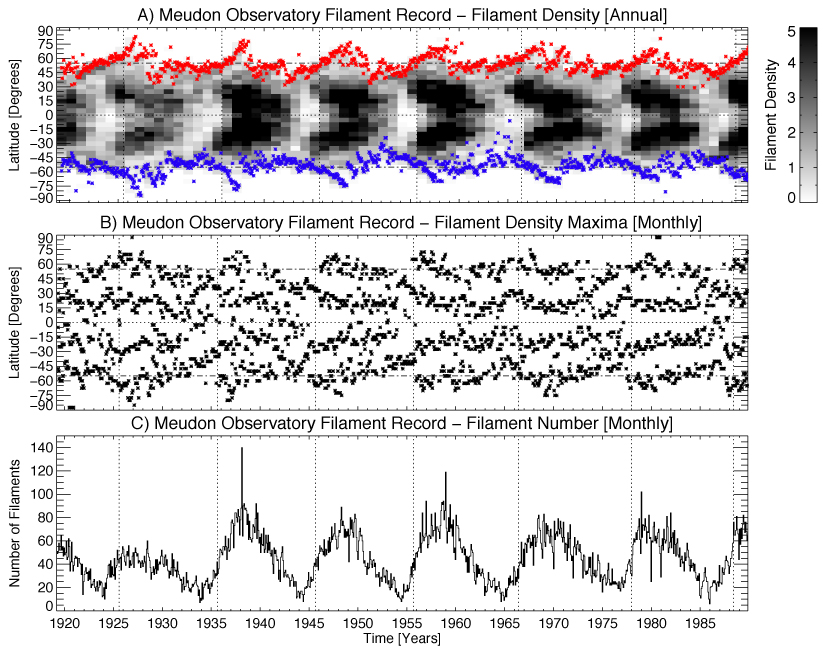}
\caption{Evolution of disk filaments on a monthly timescale, 1919--1989, provided by the Meudon Astrophysical Observatory \citep{1975SoPh...44..225H}. From top to bottom we show the monthly filament density (in 5\degree\ latitude bins) versus latitude. Included in the panel are the positions of the highest latitude filament in the northern (red~\textbullet) and southern (blue~\textbullet) hemisphere. The black~\textbullet~symbols of the middle panel indicate the latitude of the monthly filament density maxima. The lower panel shows the monthly average number of filaments. In each panel the vertical dotted lines are the times of the magnetic (Hale) cycle terminations \citep{2019arXiv190109083M}.}
\label{fig:f4}
\end{figure}

Similar to the work of Bocchino, \cite{1975SoPh...44..225H} presented diagnostics of the systematic evolution of filaments as observed continuously from the Paris Observatory at Meudon from 1919 until 1973. Figure~\ref{fig:f4} presents an analogous figure to Figure~\ref{fig:f3} demonstrates the Meudon monthly filament density, position of the highest latitude filaments present, the maxima of the filament density and the monthly number of filaments present on the solar disk. From this longer time series the persistent, overlapping, pattern in the filament density maxima is clear.

\begin{figure}[!ht]
\centering
\includegraphics[width=\linewidth]{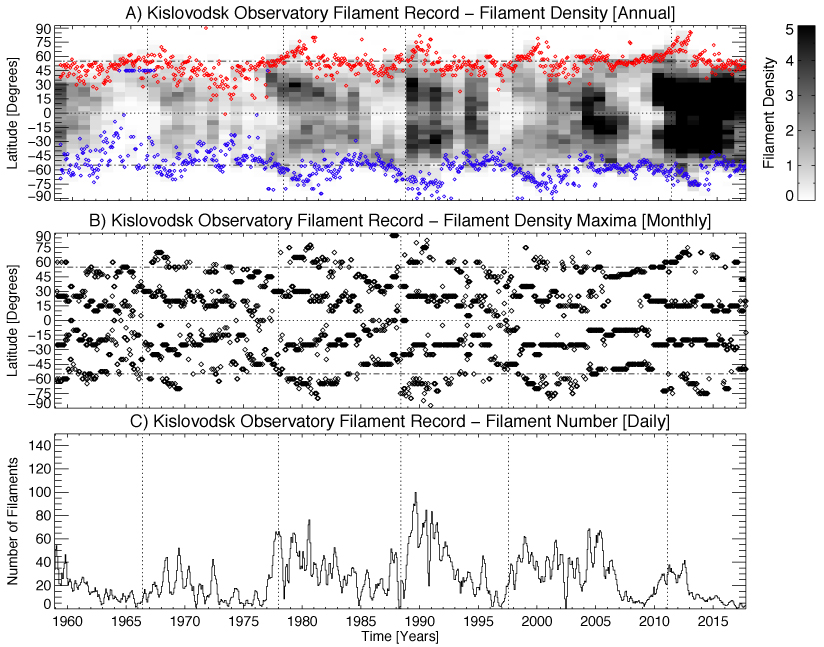}
\caption{Evolution of disk filaments on a monthly timescale, 1959--2017, provided by the Kislovodsk Solar Mountain Astronomical Station \citep{2016SoPh..291.1115T}. From top to bottom we show the monthly filament density (in 5\degree\ latitude bins) versus latitude. Included in the panel are the positions of the highest latitude filament in the northern (red~$\triangle$) and southern (blue~$\triangle$) hemisphere. The black~$\triangle$~symbols of the middle panel indicate the latitude of the monthly filament density maxima. The lower panel shows the monthly average number of filaments. In each panel the vertical dotted lines are the times of the magnetic (Hale) cycle terminations \citep{2019arXiv190109083M}.}
\label{fig:f5}
\end{figure}

The Kislovodsk Mountain Astronomical Station of the Russian Academy of Sciences has a filament record of comparable length of that from Meudon, and is extended to the present day \citep{2016SoPh..291.1115T}. Analogous to Figure~\ref{fig:f3} and Figure~\ref{fig:f4} we present the record of filaments catalogued at Kislovodsk from 1959 to 2018 in Figure~\ref{fig:f5}. Where we again show (from top to bottom), the monthly filament density, position of the highest latitude filaments present, the maxima of the filament density and the monthly number of filaments present on the solar disk. The persistent, overlapping, pattern in the filament density maxima is also clearly visible in this dataset. 

We note that the departure from around 55\degree\ is synchronized (in both solar hemispheres) with the terminator events at the solar equator \citep{2019arXiv190109083M} as indicated by the dashed vertical lines in the plot. These ``rush to the poles'' (RTTP) events mark the passage of the (global-scale) highest latitude magnetic neutral line as part of the polar magnetic reversal \citep{1961ApJ...133..572B, 1989SoPh..119..323S}. 

\subsection{Large Scale Coronal Evolution}
The Fe~{\sc xiv} 5303\AA{} coronal green line is most sensitive to plasma temperatures of around 2MK and is a well-known tracer of large-scale (coronal) magnetic morphology, and hence to the underlying magnetic field \citep{1997ApJ...485..419W}. 

\begin{figure}[!ht]
\centering
\includegraphics[width=0.75\linewidth]{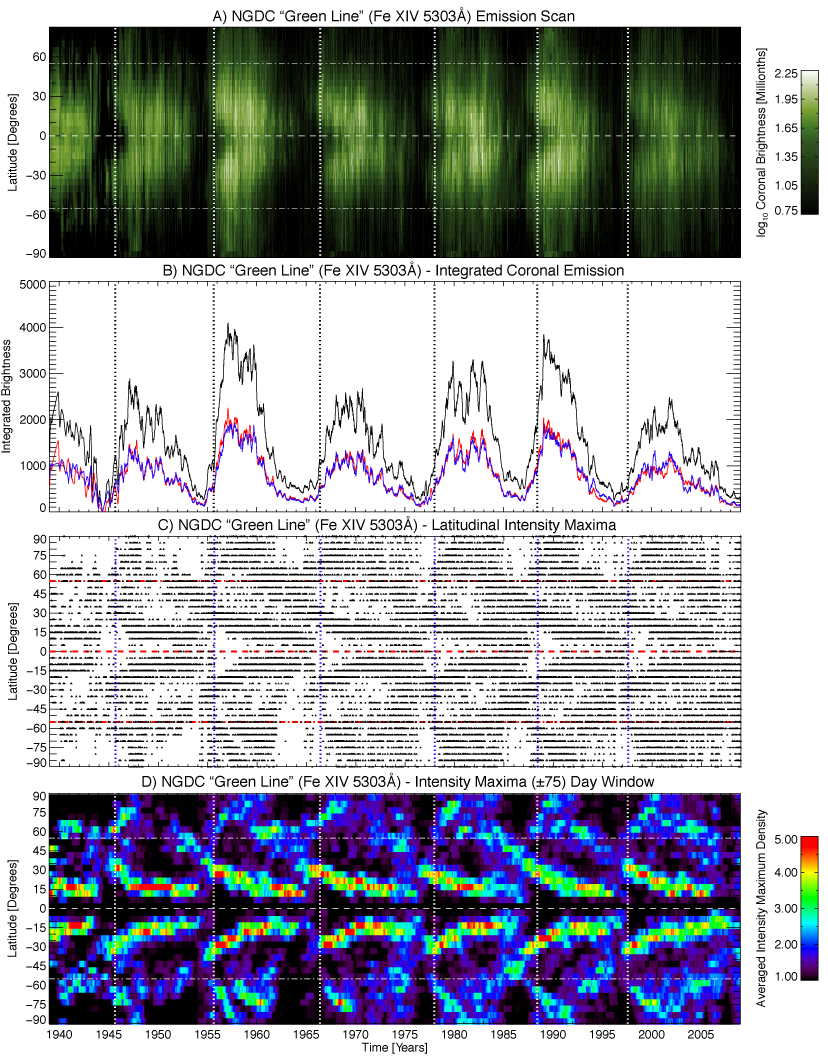}
\caption{The daily 5303\AA{} ``green line'' coronal scans from 1939 to 2009. From top to bottom we show the daily green line intensity scan (in 5\degree\ latitude bins) versus latitude. The total integrated annular coronal emission and that of each hemisphere (red \-- north; blue \-- south). The black dots of the third panel indicate the latitude of the intensity maxima identified on each daily scan. The lowest panel shows the aggregation of those maxima using a 150-day running window over the data. In each panel the vertical dotted lines are the times of the magnetic (Hale) cycle terminations \citep{2019arXiv190109083M}.}
\label{fig:f7}
\end{figure}

Originally observed during the 1869 total solar eclipse \citep{1937JRASC..31..337C}, the green line became the staple observable of the Sun's corona during eclipses and then through the development of the coronagraph \citep{1930BuAst...6..305L, 1939MNRAS..99..580L}, the means by which we still observe the corona on a daily basis. Routine observation of the green line has proceeded since the installation of Lyot's coronagraphs and their descendants. \textcolor{blue}{A repository of the daily green line scans since 1939 can be found \href{https://www.ngdc.noaa.gov/stp/solar/corona.html}{here} and is presented in Panel~A of Figure\ref{fig:f7} and see above for a description of the observational method.} 

\subsubsection{Intensity: Maximum Finding}
Referring back to Figure~\ref{fig:fA} and the discussion above we will study the position angles of the intensity maxima in the daily green line scans. We will explicitly employ the method of Altrock in the following analysis \citep[e.g.,][]{1988sscd.conf..414A, Wil88, 1997SoPh..170..411A, 2008ASPC..383..335A, Tap13, Alta14}. At the highest level Altrock's method identifies the intensity maxima in a daily scan as a function of position angle. For the daily green line record these daily maxima are represented in Panel~C of Figure~\ref{fig:f7}. To build an image of the maxima and their progression we then compile an aggregation of the maxima as a function of latitude with a rolling 150-day window---the resulting image is shown in Panel~D of Figure~\ref{fig:f7} where the color scale provides insight into the density of the intensity maxima. We can visually compare this analysis to that of Altrock's previous analyses and confirm that the same patterns exist and are prevalent. We note the presence of a similar pattern to those in the filament density maxima, featuring a lengthy overlapping pattern and an episodic migration of coronal emission to the poles (the RTTP).

\begin{figure}[!ht]
\centering
\includegraphics[width=0.75\linewidth]{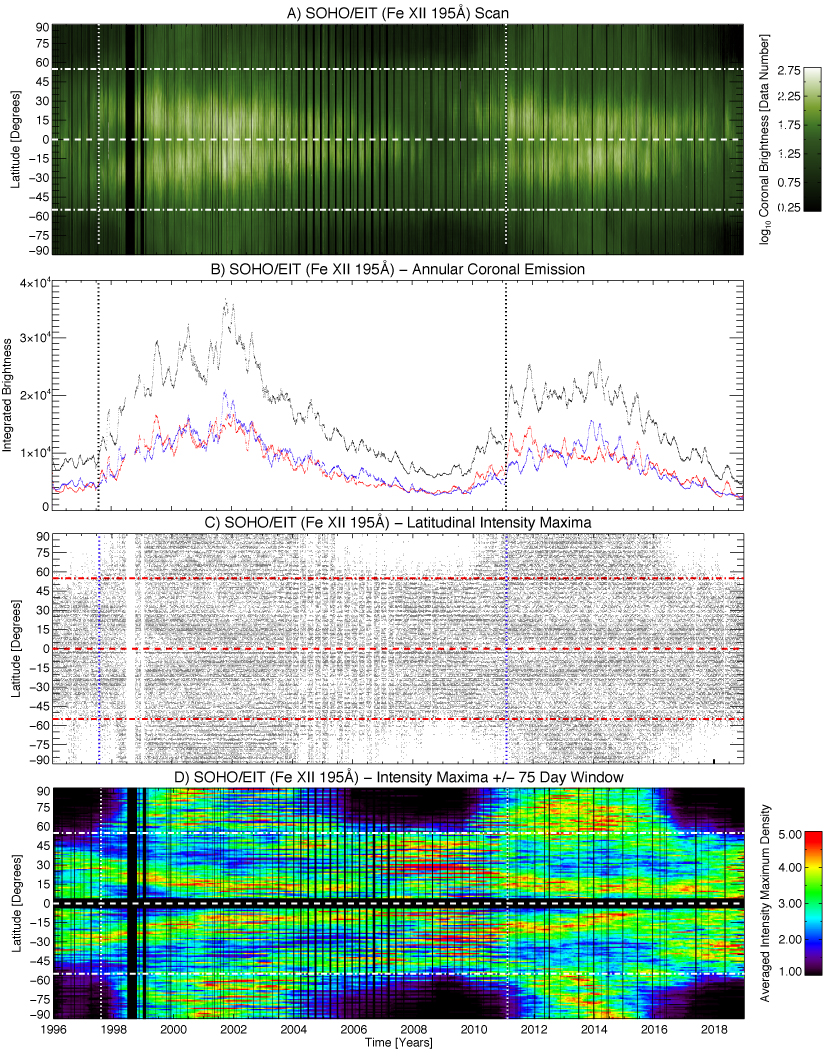}
\caption{The daily 195\AA{} \soho/EIT coronal scans from 1996 to 2019. From top to bottom we show the daily intensity scan (in 5\degree\ latitude bins) versus latitude. The total integrated annular coronal emission and that of each hemisphere (red \-- north; blue \-- south). The black dots of the third panel indicate the latitude of the intensity maxima identified on each daily scan. The lowest panel shows the aggregation of those maxima using a 150-day running window over the data. In each panel the vertical dotted lines are the times of the magnetic (Hale) cycle terminations \citep{2019arXiv190109083M}.}
\label{fig:f8}
\end{figure}

\subsection{Contemporary Picture of Coronal Morphology}
It is straightforward to extend the analysis of Altrock to the fleet of contemporary EUV imagers in space (see, e.g., Figure~\ref{fig:fA} and the discussion above). In the following figures we replicate the maximum finding analysis on broadband emission from the 195 and 193\AA{} channels of \soho{}/EIT, \stereo/EUVI, and \sdo/AIA imagers. 

Figure~\ref{fig:f8} replicates the analysis presented in Figure~\ref{fig:f7} identically, but applied to the daily synoptic image series of \soho/EIT using the image taken nearest to midnight UT as the baseline. Given the nearly 24 years of coronal observation from \soho{} it is possible to clearly see the overlapping structure below 55\degree\ latitude and the polar excursions that start in 1998 and 2011 that are associated with the magnetic (Hale) cycle terminations \citep{2019arXiv190109083M}.

\begin{figure}[!ht]
\centering
\includegraphics[width=0.95\linewidth]{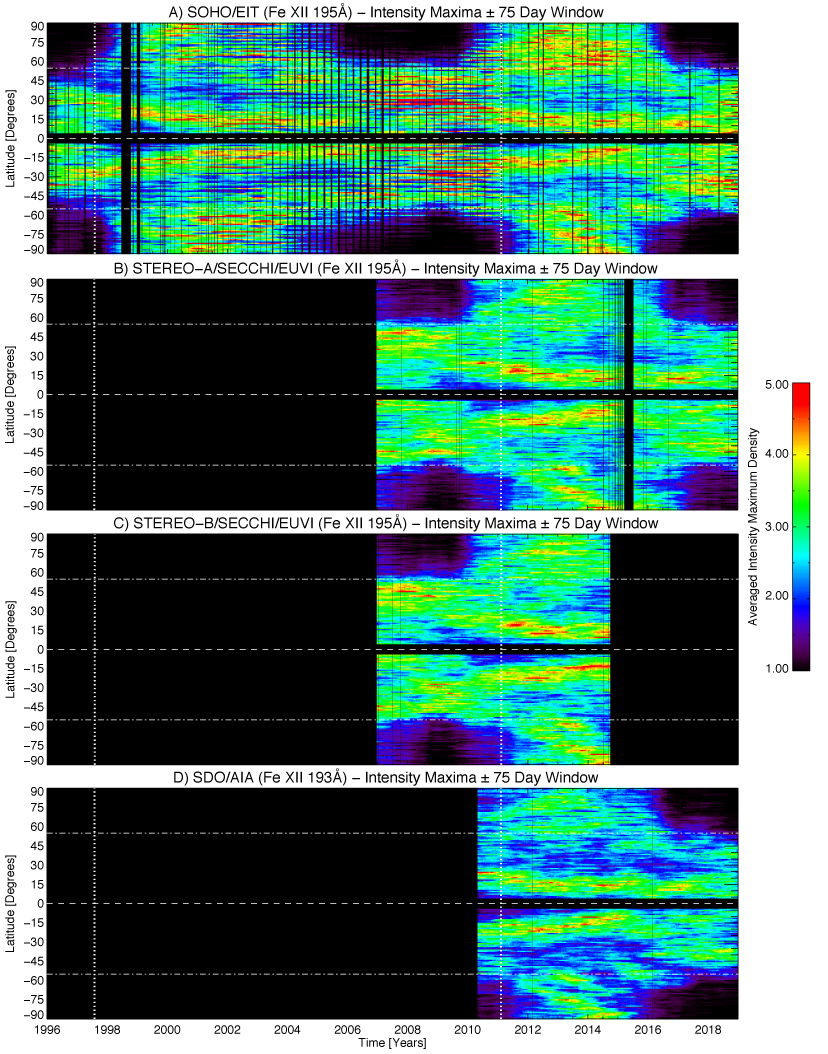}
\caption{A compendium of the daily EUV intensity scans from 1996 to 2019. Again, replicating Altrock's intensity maximum finding approach, we show the intensity maxima distributions from the fleet of space-borne EUV imagers. From top to bottom we show \soho{}/EIT (Panel~D of Figure~\ref{fig:f8}), \stereo/EUVI ``Ahead'', ``Behind'', and \sdo/AIA. In each panel the vertical dotted lines are the times of the magnetic (Hale) cycle terminations \citep{2019arXiv190109083M} while the horizontal dot-dashed lines signify 55\degree\ latitude}
\label{fig:f10}
\end{figure}

Unsurprisingly, as shown in Figure~\ref{fig:f10}, the four EUV imagers in the same wavelength range reinforce one another with the highest spatial resolution imager, SDO/AIA, providing the cleanest picture of the multi-banded emission structure. In that image one can clearly see the emergence and progression of four bands, two at low and two at mid latitudes, the lower being immediately responsible for sunspot cycle 24 and the the higher ultimately being responsible for sunspot cycle 25.

\newpage

\section{Results}
In the previous section we have observed a host of chromospheric and coronal morphological features of the global-scale magnetic field that exhibit what appears to be the same pattern. In the following analysis we will utilize the Hale cycle terminators to illustrate the cyclic nature of these patterns over the entire course of observational imaging. 

\subsection{Using Terminators as a Fiducial Clock}
As we have discussed above the termination of Hale magnetic cycles at the Sun's equator triggers the rapid growth of the sunspot butterfly pattern and the initiation of the rush to the poles over all of the 14 solar cycles observed in detail since 1860. In the following analysis we will use the terminator times recorded by \cite{2019arXiv190109083M}, and verified algorithmically by \cite{2020SoPh..295...36L}, to perform a Superposed Epoch Analysis \citep[SEA;][]{1913RSPTA.212...75C} on all of the observational diagnostics shown above using the terminators as the key time, or fiducial clock for the SEA.

Superposed Epoch Analysis is a statistical tool used to detect periodicities within time series, or as we will employ it, to reveal recurrence of patterns in, and between, time series. The essence of the method is to first define each occurrence of an event in one data sequence as the key time---in our case the terminators, then extract subsets of data from the other sequence within some time range near each key time---for our analysis we choose a period of $\pm$11-years, and finally superpose all extracted subsets from the time series (with key times for all subsets synchronized) by adding them. This approach illustrates underlying correlation when the superposed signals are in phase with one another, reinforcing the additive sum, or adding to smear them out in the event that the additive sums are not in phase. 

\begin{figure}[!ht]
\centering
\includegraphics[width=\linewidth]{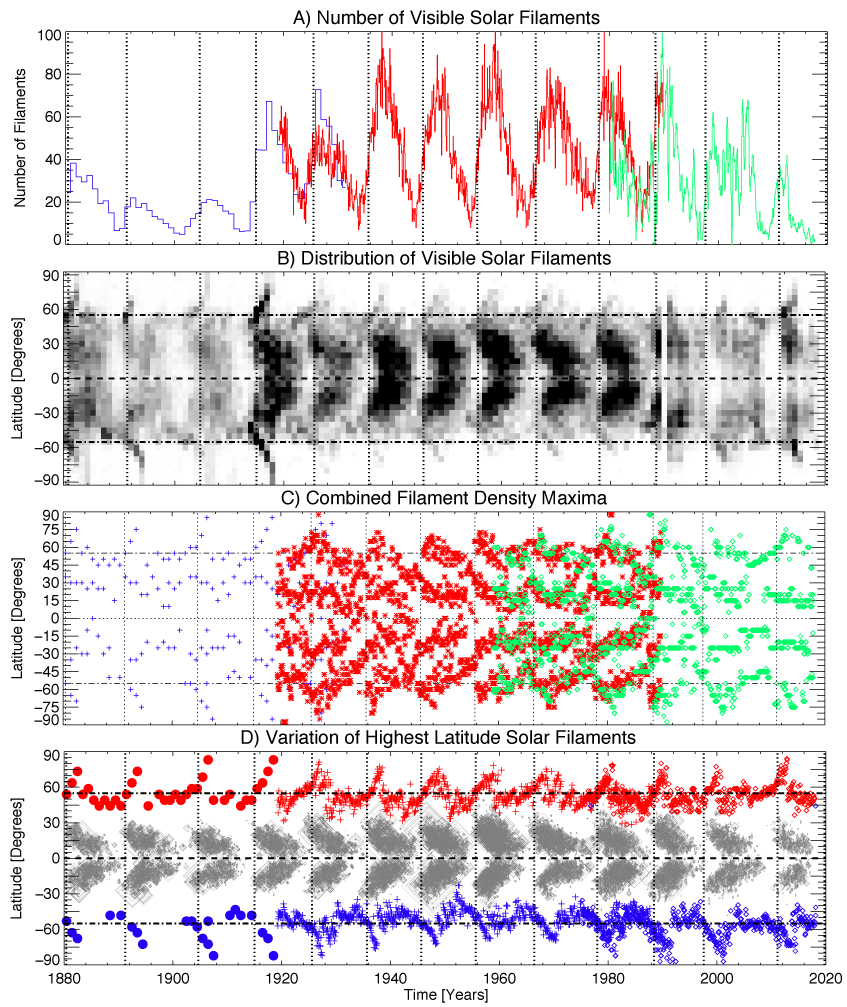}
\caption{A unified record of solar filament evolution from 1880 to 2017. Combining the data and analyses shown in Figs.~\ref{fig:f3},~\ref{fig:f4} and~\ref{fig:f5} from the three observatories we see almost 140 years of filament evolution. Panel~A shows the filament number, Panel~B shows the filament density, Panel~C shows the maxima of the filament density as a function of timestep and latitude, and finally Panel~D shows the highest latitude filaments in each hemisphere (red \-- north; blue \-- south) in comparison to the sunspot butterfly diagram (in grayscale) for comparison to Figure~\ref{fig:f1}.}
\label{fig:f6}
\end{figure}

\subsection{Superposed Epoch Analysis on The Spectrum of Solar Activity}
The figures above have illustrated the variability of a number of solar features, each tied to the global-scale magnetic field in patterns that emerge at high latitudes and migrate over a period of years to the equator. These global patterns do not exist on the Sun separately, they overlap in time and for almost all of the time there are two such patterns visible. Given the analysis of \cite{2019arXiv190109083M} we wish to explore the spatio-temporal relationship of these patterns further using the SEA method and the terminator times as the key time. Figure~\ref{fig:f6} provides the starting point for the 140 year composite of solar filaments, Figure~\ref{fig:f7} for the daily green line scans of the corona, Figure~\ref{fig:f8} for the \soho{}/EIT 195\AA{} coronal intensity scans, and Figure~\ref{fig:f10} for the \stereo{} and \sdo{} records before comparing the results with the sunspot butterfly diagram of Figure~\ref{fig:f1}.

As an illustrative example of SEA applied to this data let us consider the SEA decomposition of Figure~\ref{fig:f8}D in Figure~\ref{fig:f12}. The \soho{}/EIT record shown contains the 1997 and 2011 terminators; extrapolating out in time would place the cycle 24 terminator in mid 2020 \citep{2020SoPh..295...36L} such that we have three terminators and have four epochs of 22 years (plus and minus 11 years from each terminator) that we want to compare and contrast. Panel~A of Figure~\ref{fig:f12} shows the complete \soho{}/EIT dataset with the four epochs shown as red, blue, green and orange rectangles that are extracted as the remaining panels of the figure. 

The reader will immediately see the strong visual correspondence of the patterns present in each of the short epoch. Figure~\ref{fig:f13} shows the mean of the added (non-zero) signal from panels B through E of Figure~\ref{fig:f12}. It illustrates the pattern the eye is drawn to, one of two temporally overlapping chevrons between plus and minus 55\degree\ latitude with correspondingly aligned rush to the poles features. If these patterns were out of phase with the terminators their appearance would be significantly smeared. Their strong correlation of the rush to the pole features with the terminator was previously noted \citep{2019arXiv190109083M}.

\begin{figure}[!ht]
\centering
\includegraphics[width=0.85\linewidth]{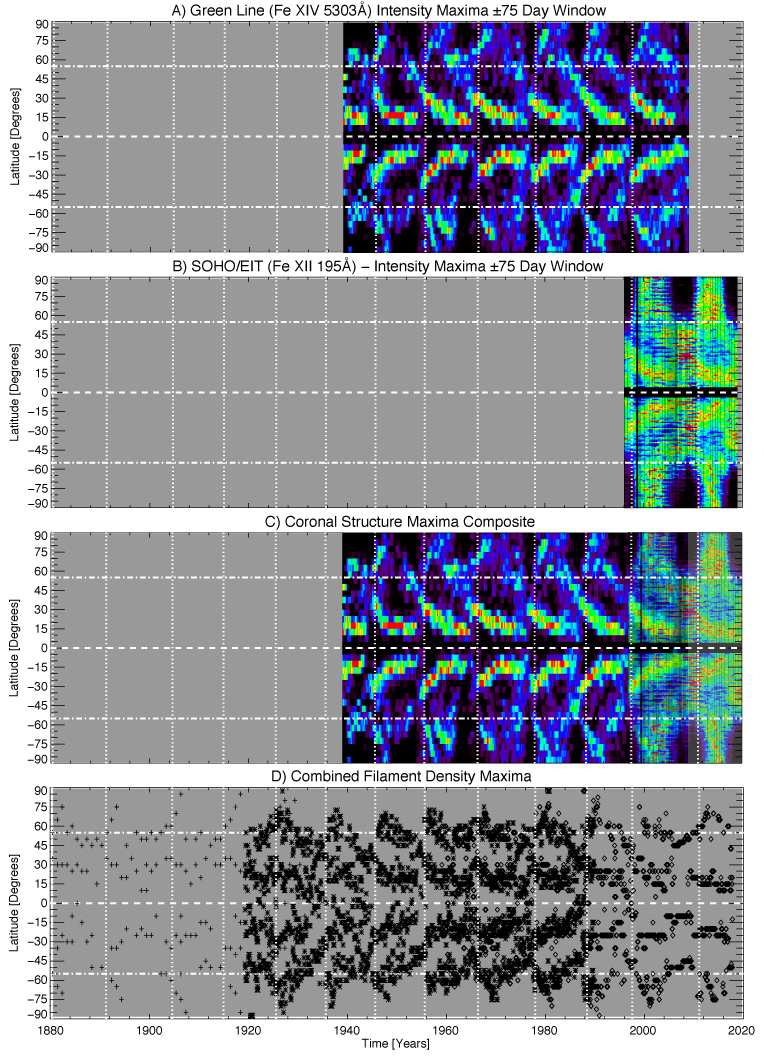}
\caption{Comparing the coronal intensity and filament density maxima versus latitude over the past 140 years. From top to bottom we compare Figure~\ref{fig:f7}D with Figure~\ref{fig:f8}D, their merged signature, with the filament density maxima of Figure~\ref{fig:f6}C. In each panel the white vertical dotted lines are the times of the magnetic (Hale) cycle terminations \citep{2019arXiv190109083M}, the horizontal dashed line indicates the equator while the horizontal dot-dashed lines signify 55\degree\ latitude in each hemisphere.}
\label{fig:f11}
\end{figure}

While the \soho{}/EIT example is illustrative of the SEA method's application to these data the relatively short span of the time series is not fully revealing of the recurrent underlying pattern in these data. Figure~\ref{fig:f14} sees the application of the SEA method to the data in Figure~\ref{fig:f11} with the panels, from top to bottom, showing the density functions of the filament density maxima, the green line coronal intensity maxima, with the EUV record of Figure~\ref{fig:f13}. To put these in context we perform the SEA analysis of the sunspot butterfly data shown in Figure~\ref{fig:f1}B. We note that the online edition of the journal has an animation of the SEA analysis for each of panels B through E to illustrate the application of the method in each case. 

\begin{figure}[!ht]
\centering
\includegraphics[width=0.75\linewidth]{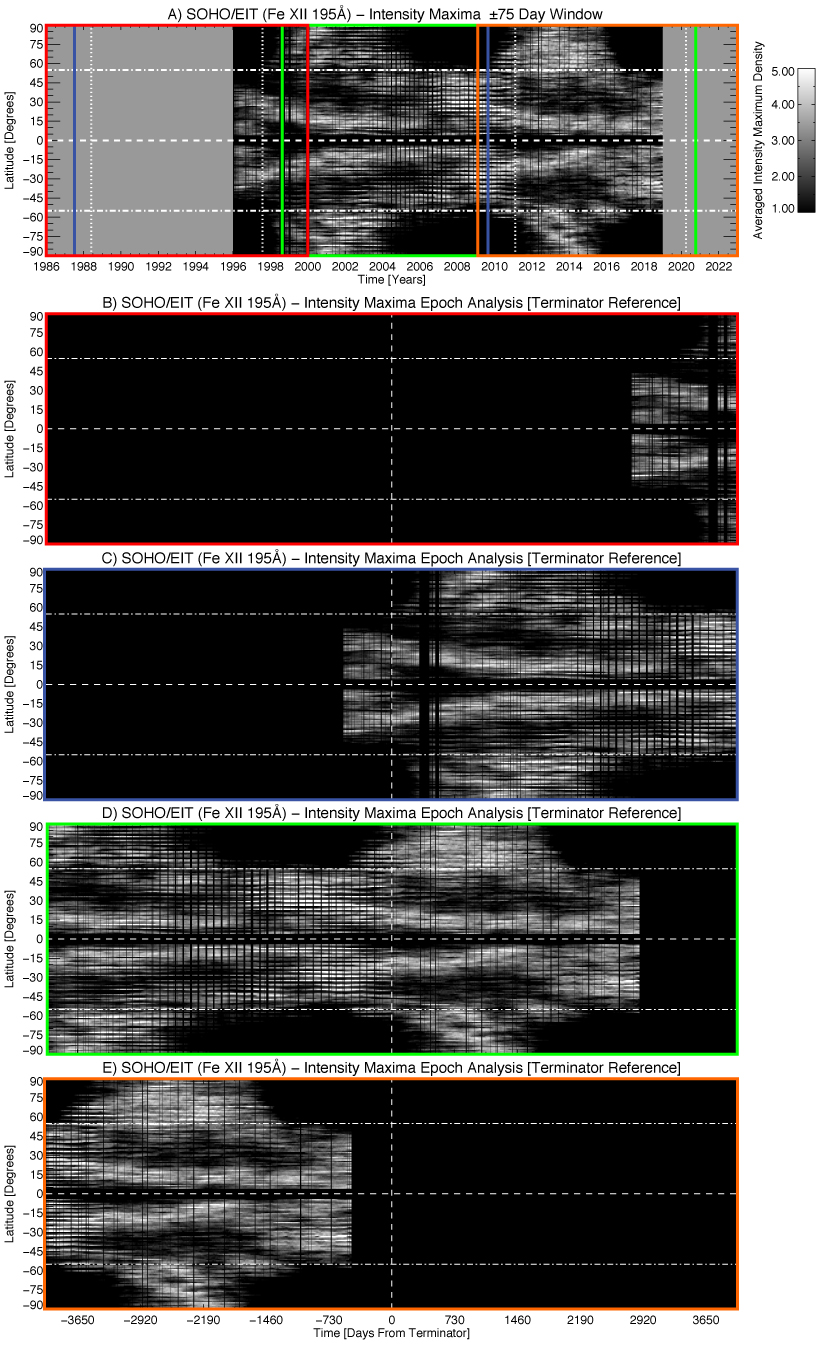}
\caption{Illustrating the Superposed Epoch Analysis as applied to \soho{}/EIT coronal intensity maxima. From top to bottom we show the complete dataset (see, Figure~\ref{fig:f8}D) with the key terminator times as vertical dashed white lines. Also shown are the 22-year long epochs related to those terminators that form the basis for panels B through E. In each panel the white vertical dotted lines are the times of the magnetic (Hale) cycle terminations \citep{2019arXiv190109083M}, the horizontal dashed line indicates the equator while the horizontal dot-dashed lines signify 55\degree\ latitude in each hemisphere.}
\label{fig:f12}
\end{figure}

\begin{figure}[!ht]
\centering
\includegraphics[width=\linewidth]{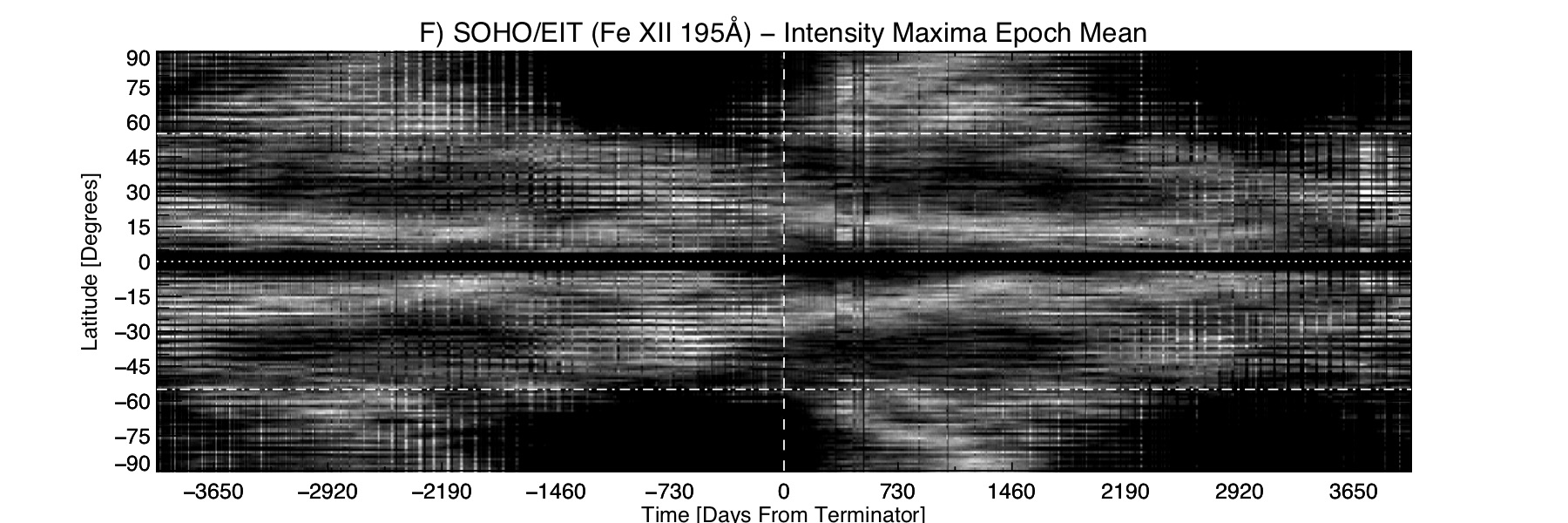}
\caption{The SEA-derived mean \soho{}/EIT coronal intensity maxima pattern as constructed from panels B through E of Figure~\ref{fig:f12}. The horizontal dashed line indicates the equator while the horizontal dot-dashed lines signify 55\degree\ latitude in each hemisphere.}
\label{fig:f13}
\end{figure}

\subsection{Strongly Cyclic Behavior}
\begin{figure}[!ht]
\centering
\includegraphics[width=0.9\linewidth]{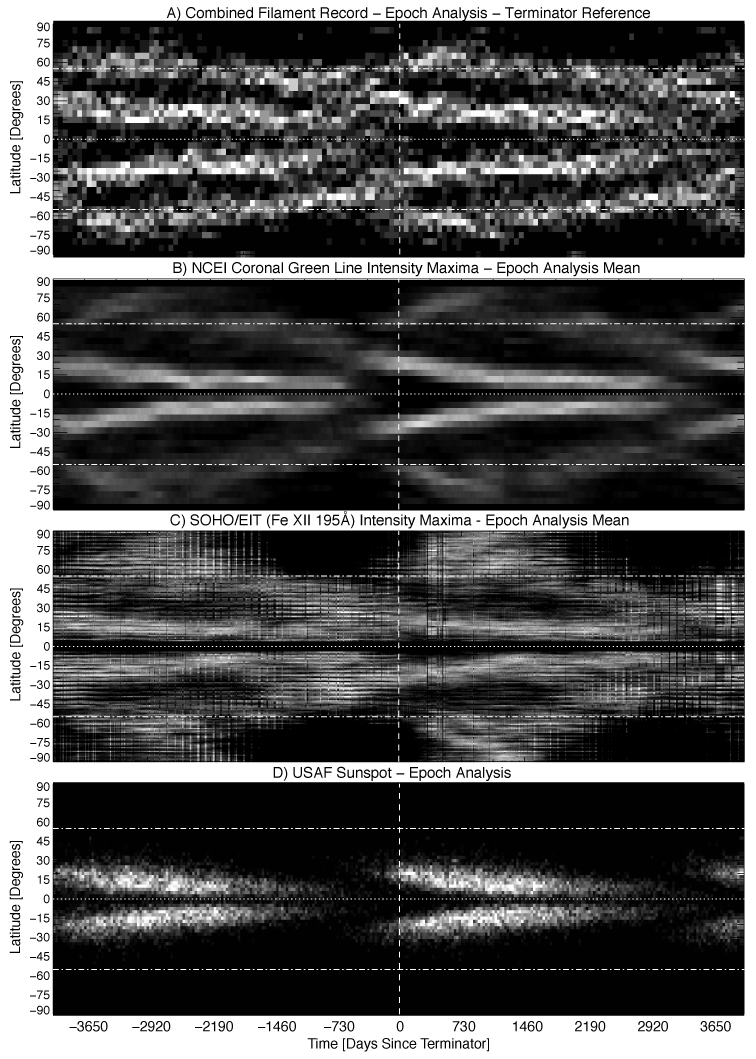}
\caption{Comparing the SEA-derived mean patterns of filaments, coronal green line, EUV coronal intensity maxima, and sunspots. The vertical dashed line indicates a time of zero. The horizontal dashed line indicates the equator while the horizontal dot-dashed lines signify 55\degree\ latitude in each hemisphere. The online edition of the journal has an animation to illustrate the SEA analysis for each of panels presented here.}
\label{fig:f14}
\end{figure}

This SEA analyses applied to the remaining datasets is clearly demonstrated in Figure~\ref{fig:f14}. The 140 years of filament data, when aligned using the terminators and explored by SEA, reveals a recurrent overlapping pattern of activity that starts around 55\degree\ latitude in each hemisphere. Branches of that recurrent pattern move poleward and equatorward---the latter taking considerably longer, almost the entire 22 year period, to migrate from 55\degree\ to the equator. That pattern is mirrored in the other panels until we reach the terminator-keyed Sunspot pattern where we see that the mean sunspot pattern does indeed collapse through the SEA, but only fills half of the timeframe of the overall 22-year Hale cycle pattern and also only a subset of the latitudinal progression---this is precisely the hypothesis established by M2014---the sunspot butterfly must be a derivative of the interplay between the magnetic bands of the 22-year Hale cycle. This figure unequivocally demonstrates that relationship and cements the link.

This recurrent spatio-temporal pattern, approximately 22-years in duration is manifest from 1960 to the present in a host of atmospheric diagnostics, each of which is a tracer of the Sun's global-scale magnetic field. These analyses are also consistent with a systematic analysis of limb prominences over 10 Solar Cycles using images from Kodaikanal, Meudon, and Kanzelh\"ohe Observatories \citep{2020E&SS....700666C}. The pattern observed is one and the same with that of M2014 that is illustrated in Figure~\ref{fig:f2}. The SEA analysis of the sunspot butterfly (Figure~\ref{fig:f1}) and its well-contained regularity when using the terminator as a fiducial clock, in concert with Figs.~\ref{fig:f2} and~\ref{fig:f14}, only highlights that the number of sunspots produced by the Sun, which is at best quasi-periodic \citep{2010LRSP....7....1H}, is a variant of the spatio-temporal interaction of the underlying Hale magnetic systems, as the hypothesis of M2014 lays out. 

\subsection{Connecting to the Interior}
\begin{figure}[!ht]
\centering
\includegraphics[width=0.8\linewidth]{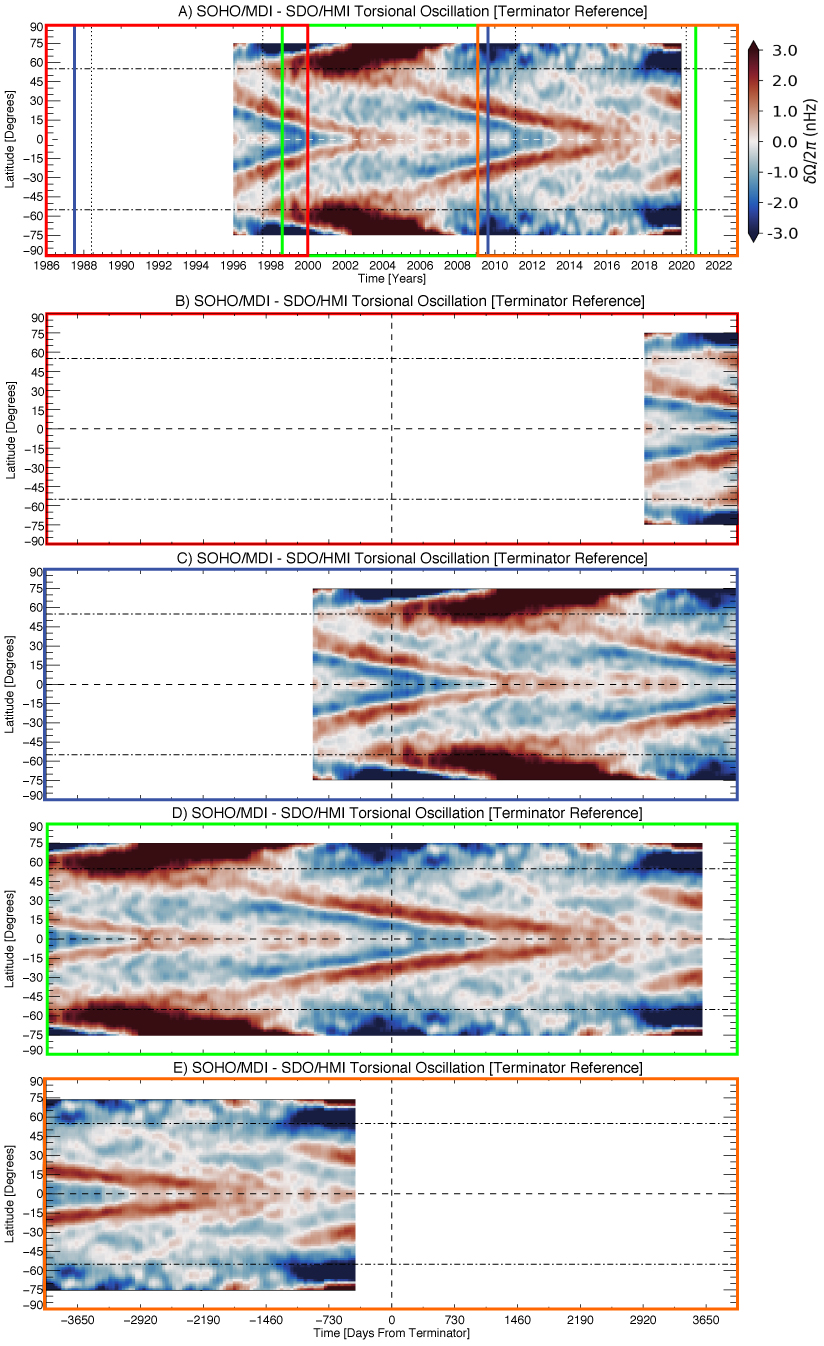}
\caption{Illustrating the Superposed Epoch Analysis as applied to the solar interior's Torsional Oscillation. Compare to Fig~\ref{fig:f12} From top to bottom we show the complete dataset (see, Figure~\ref{fig:f8}D) with the key terminator times as vertical dashed ]\textcolor{blue}{black} lines. Also shown are the 22-year long epochs related to those terminators that form the basis for panels B through E. In each panel the white vertical dotted lines are the times of the magnetic (Hale) cycle terminations \citep{2019arXiv190109083M}, the horizontal dashed line indicates the equator while the horizontal dot-dashed lines signify 55\degree\ latitude in each hemisphere. The data shown were provided by Rachel Howe.}
\label{fig:f15}
\end{figure}

All of the proxies of global-scale magnetism that we have explored above are chromospheric or coronal in nature. M2014 illustrated that the helioseismically inferred ``torsional oscillation'' \citep[e.g.,][]{Lab82,Howe09} exhibited the same spatio-temporal pattern as the EUV brightpoints. The torsional oscillation is a latitudinal system of alternating faster and slower bands (relative to the mean differential rotation at each \textcolor{blue}{latitude}. \cite{Wil88} demonstrated that coronal green line maxima and torsional oscillation also shared the same pattern. Therefore it is not a surprise that applying the terminator-key SEA to the torsional oscillation pattern (Figure~\ref{fig:f15}) follows those of Figure~\ref{fig:f14} and so our recurrent global-scale magnetic field pattern is also manifesting in the solar interior---its almost certain origin.

\section{Discussion}
As stated above and illustrated in Figure~\ref{fig:f14} the superposed epoch analysis of these data yields a recurrent, consistent, coherent pattern of global scale evolution in the interior, chromosphere, and corona over the entire 140 years of photographic observation of the Sun. 
This pattern is a manifestation of the Sun's 22-year Hale magnetic cycle. It is not too much to presume that the solar interior and its evolution is directly driving the outer atmosphere and the patterns that we observe there. It must also follow that, especially given the subset nature of the sunspot butterfly pattern relative to the global-scale spatio-temporal pattern, the modulation in number of magnetic spots and active regions has a common root.

\textcolor{blue}{The repeating global-scale magnetic pattern has several key properties:
\begin{itemize}
\item[{\em i\/}]{Let us assume that the terminator is the starting point: originating around 55\degree\ a branch departs to the pole over the course of about 3.5 years at an average speed of 10\degree\ per year in each hemisphere (the ``Rush to the Poles'');}
\item[{\em ii\/}]{a second branch lingers around 55\degree\ until a time commensurate with that at which the polar branch completes its journey before it starts migrating towards the equator;}
\item[{\em iii\/}]{the migration to the equator takes about 19 years to complete at an average speed of about 2.9\degree\ per year;} 
\item[{\em iv\/}]{the initial speed of the equatorial migration is faster than the later phases; approximately every 11 years a new high latitude branch forms at 55\degree\ that has opposite magnetic polarity;}
\item[{\em v\/}]{the features observed are not necessarily symmetric across the equator.}
\end{itemize}
}

\textcolor{blue}{Point {\em iii\/} above means that, for about than 75\% of the 22 year period, there are four oppositely signed magnetic bands between 55\degree\ and the equator.} The exception being the rise phase of the solar cycle from the terminator commensurate with the high latitude ``Rush to the Poles.'' The observed average behavior over 140 years will prompt a revision to the linear approximations that are used to construct the ``band-o-gram'' of M2014 (Figure~\ref{fig:f2}D).

As mentioned above, similar patterns have been discussed in the literature, prevalently in the mid to late 1980s as the ``extended solar cycle,'' but seldom since. The remainder of the discussion leans heavily on the pioneering work of Altrock, Harvey, Martin, McIntosh (Patrick), Snodgrass, Wilson, and others that was captured in a 1988 volume of Solar Physics (Vol.~117, Issue 2) following a workshop dedicated to the subject the previous year, and in Wilson's essential monograph \citep{1994ssac.book.....W}.

\cite{1994ssac.book.....W} could not reconcile conventional dynamo models \citep{2010LRSP....7....3C} with the presence of the extended cycle pattern. As we have mentioned above, the sunspot pattern is subset of the Hale cycle band progression that makes up the extended cycle. Therefore, it would appear that the recurrent wave-like pattern of the Hale cycle plays an important role in the shape of the sunspot pattern and the number of spots that it produces. To highlight this, we consider the phenomenon of solar/sunspot minimum. There are four oppositely signed magnetic bands of the Hale Cycle within 35\degree\ of the equator---a simple explanation for the lack of spot production in this time (recalling that small-scale magnetic features like the EUV BPs in Figure~\ref{fig:f2} do not stop) is that there is insufficient free magnetic flux to buoyantly emerge a large concentration like a spot. By extension, the initial appearance of sunspots from the band seems to be more dependent on where the mid-latitude band {\em is\/} when the terminator occurs than anything else, although on average (by Figure~\ref{fig:f14}) is about 25-30\degree. If the Hale pattern near the equator drifts too far to the right (i.e., in time, or put another way, slows its equatorward progression), it starts to impact the growth time of the flux emergence on the mid-latitude band---this appears to have significantly impacted the amplitude of sunspot cycle 24 although the underlying reason for the drift in time (slower latitudinal motions) is at the present time unclear \citep{2020SoPh..295...36L}.

With the recurrence of the Hale Cycle (Extended Solar Cycle) pattern we should relaunch the discussion of \cite{1987SoPh..110...35S} and \cite{1987Natur.328..696S} as to the nature of the circulatory patterns in the solar interior and their intersection with the convective pattern. The prevalence of 55\degree\ latitude in this puzzle {\em throughout\/} the entire observational record (Figure~\ref{fig:f6}) highlights the presence of a strong, possibly geometric, demarcation between the polar and equatorial zones. We note that this heliographic latitude marks approximately the photospheric intersection of a cylinder tangent to the base of the convection zone at the equator, a location where hydrodynamic and magnetohydrodynamic simulations show a change in convective pattern from rotation-axis-aligned elongated ``banana cells'' below, to smaller convective cells above \citep[e.g.,][]{2009AnRFM..41..317M,2013ApJ...762...73N,2014ARA&A..52..251C}.
Detailed measurement of the high latitude flow structure, particularly from the ecliptic plane, is challenging, and these latitudes are approaching the limits of helioseismic analysis with line-of-sight velocities, but the detected features tell a story while we wait for the observations of Solar Orbiter over the next decade. The creation and recurrence of the polar crown filament and the paucity of EUV BPs above 55\degree, are not ambiguous (and {\em not limited\/} by the viewing angle from the ecliptic plane). The former points at persistent shear flows and magnetic neutral lines, the latter in differences in the underlying flux emergence or convective structure \citep{2014ApJ...784L..32M}. 

At latitudes lower than 55\degree\ there is a clear migration of a magnetic system towards the equator and that magnetic system maps one-to-one with the underlying torsional oscillation pattern (Figs.~\ref{fig:f12} and~\ref{fig:f15}). This pattern will lie beneath, above, or thread the multi-scale granulation. Of these, the threading or deep options are most likely---it is hard to believe that a pattern like this is formed and maintained in the highly turbulent convective surface regions for 140 years continuously. Threading the convective pattern cannot be overruled based on the helioseismic determination of the torsional oscillation pattern. The deep-rooted option was proposed by 
\cite{1987Natur.328..696S}. 
We would tend to support the latter too: the torsional oscillation must be a tracer of significantly organized large-scale magnetic systems and the thermodynamic impact they have on the flow at that scale field \citep[e.g.,][]{2006ApJ...647..662R,2019shin.confE.199M} and this is embedded in the deeper parts of the convection zone. We also cannot escape the suggestion that the global-scale magnetic fields present are of substantial magnitude: strong enough to exhibit the coupled global signature of the terminator events going back over a century \citep[e.g.,][]{2019NatSR...9.2035D,2019arXiv190109083M}, and to modify the circulation of the interior plasma as exhibited by the torsional oscillation. 

\section{Conclusion}
The analysis above indicates that the Hale magnetic cycle, as highlighted by the terminator-keyed superposed epoch analysis, is strongly recurrent throughout the entire observational record studied, some 140 years. The subset nature of the sunspot butterfly pattern to the underlying Hale cycle strongly suggests that the production of sunspots is not the fundamental feature of the cycle, and is merely a product of the Hale magnetic cycle and that the modulation in amplitude of the sunspot cycle is critically dependent on how Hale cycles magnetically interact in latitude and time. The Hale cycle pattern highlights the importance of 55\degree\ latitude in the evolution, and possible production, of solar magnetism. 

These results presented above highlight the scientific potential of synoptic observing and archiving of solar observations---the present disposition of such programs is, at best, poor. The potential of the diagnostics used above to identify and unambiguously track the presence of the Hale cycle throughout the solar atmosphere, as an essential boundary condition for future predictive dynamo models, emphasizes the need to reconstitute a national ground-based synoptic program. 

\section*{Acknowledgements}
This paper is dedicated to the memory of Donald~L.~M.~McIntosh (1939\--2021). This material is based upon work supported by the National Center for Atmospheric Research, which is a major facility sponsored by the National Science Foundation under Cooperative Agreement No.~1852977. RJL acknowledges support from NASA's Living With a Star Program. SMC acknowledges the data ingest skills of Sheryl Shapiro, the encouragement, enthusiasm and insight of Frank M. Howell. We also acknowledge the grant of Indo-US Virtual Networked Center (IUSSTF-JC-011-2016) to support the joint research on ESCs.

\section*{Conflict of Interest Statement}
The authors declare that there is no conflict of interest

\bibliographystyle{spr-mp-sola}

\end{document}